\documentclass[showpacs,twocolumn,aps,prl]{revtex4}
\usepackage{amsmath}
\usepackage{amsfonts}
\usepackage{amssymb}
\usepackage{amsthm}
\usepackage{graphicx}

\begin{document}

%----------------------------------------------------------------%
\title{Fast magnetization switching of Stoner particles:
A nonlinear dynamics picture}
%\title{Flow diagram of fast magnetization switching in phase plane}

\author{Z. Z. Sun}
%\email[To whom correspondence should be addressed. Electronic
%address: ]{phszz@ust.hk}
\affiliation{Physics Department, The Hong Kong University of
Science and Technology, Clear Water Bay, Hong Kong SAR, China}
\author{X. R. Wang}
\affiliation{Physics Department, The Hong Kong University of
Science and Technology, Clear Water Bay, Hong Kong SAR, China}

\date{\today}

\begin{abstract}
The magnetization reversal of Stoner particles is investigated
from the point of view of nonlinear dynamics within the
Landau-Lifshitz-Gilbert formulation. The following results are
obtained. 1) We clarify that the so-called Stoner-Wohlfarth (SW)
limit becomes exact when damping constant is infinitely large.
Under the limit, the magnetization moves along the steepest energy
descent path. The minimal switching field is the one at which
there is only one stable fixed point in the system. 2) For a given
magnetic anisotropy, there is a critical value for the damping
constant, above which the minimal switching field is the same as
that of the SW-limit. 3) We illustrate how fixed points and their
basins change under a field along different directions. This
change explains well why a non-parallel field gives a smaller
minimal switching field and a short switching time. 4) The field
of a ballistic magnetization reversal should be along certain
direction window in the presence of energy dissipation.
%The width of the window increases with the damping constant.
The width of the window depends on both of the damping constant
and the magnetic anisotropy. The upper and lower bounds of
the direction window increase with the damping constant.
The window width oscillates with the damping constant for a given
magnetic anisotropy. It is zero for both zero and infinite damping.
%At the zero damping, the width is zero, and the field direction
%must be perpendicular to the easy axis.
Thus, the perpendicular field configuration widely employed in
the current experiments is not the best one since the damping
constant in a real system is far from zero.
\end{abstract}
%\keywords{}
\pacs{75.60.Jk, 75.75.+a, 05.45.-a}
% 75.60.Jk Magnetization reversal mechanisms\\
%75.75.+a Magnetic properties of nanostructures \\
% 05.45.-a Nonlinear dynamics and nonlinear dynamical systems

\maketitle
%----------------------------------------------------------------%
\section{I. Introduction}
Magnetic data storage is one of the important components of
modern computers. Data input and output involve switching the
magnetization of magnetic storage cells (magnetization reversal).
The typical switching time with currently used technology is of
order of nanosecond. If one wants to have a faster computer
(modern electronic computers are working at a clock speed of
order of GHz), the conventional magnetization reversal method
shall soon (the clock speed is doubled every year in the past)
become a bottleneck. Thus fast magnetization switching shall
be of great importance for the future development of high speed
information industry.

The reversal of a magnetization can be achieved in many
different ways, and it is a very complicated issue\cite{book1}.
For example, in a bulk material, the magnetization reversal can
go through bucking and curling modes, or nucleation and domain
formation. The recent advance of technology allows us to fabricate
the magnetic nano-particles that are believed to be useful for
high density information storage\cite{Sun,Black,Woods,Zitoun}.
For a magnetic nano-particle, the magnetic moments of all atoms
are normally aligned in the same direction, creating a so-called
single magnetic domain. Such a nano-particle is usually called
a Stoner-Wohlfarth or Stoner particle. The understanding of
magnetization reversal of a single magnetic domain should be
relatively simple in comparison with that in a bulk system, but
important in nano-technologies\cite{Hillebrands}.

There are two challenging issues about magnetization reversal.
One is how to have a short reversal time, and the other is how
to make the switching field to be small. The conventional
magnetization reversal technique is to apply a magnetic field
antiparallel to the initial magnetization. A large enough
field can drive the initial state out of local minimum and at
the same time make the target state to be the global minimum.
Thus the system can roll down to the target state through
ringing effect\cite{Hiebert,Acremann2,Crawford}.
%The typically
%reversal time is order of nanoseconds at a field of teslas.
For the issue of minimal switching field, the classical result,
called Stoner-Wohlfarth (SW) limit, was given by Stoner and
and Wohlfarth\cite{Stoner}.
%classical result of minimal switching field was given by Stoner
%The idea is to make the target state to be a (global) minimum
%while the initial one is not a local minimum in the energy
%landscape. However, the system can only
%gradually dissipate its energy during a precessional motion.
%It will move around the precession axis (along the B-field)
%many times (ringing phenomenon or ringing mode)\cite
%{Hiebert,Acremann2,Crawford} before reaching the target state.
%As a result, it takes typically nanoseconds to switch a
%magnetization at a field of teslas.

Recently picoseconds magnetization switching has been observed in
experiments\cite{Back,Schumacher} by using pulse magnetic fields.
Unlike the conventional method, the magnetic field is applied
in a perpendicular direction such that the magnetization
undergoes a precession. This approach has also received
many theoretical attentions\cite{He2,Bauer,Acremann1,Miltat}.
Numerical investigations\cite{He2} showed that the switching
time can be substantially reduced because ringing
effect can be suppressed so that the magnetization will
move along a so-called ballistic trajectory\cite{Miltat}.
The precessional magnetization reversal provides not only a
shorter time but also lower switching field (well below the
SW-limit), as found in the early numerical calculations\cite{He1}.
%The switching field could be well below the SW-limit.
In the absence of energy dissipation, precessional magnetization
switching can also be investigated analytically. Analytical results
for the minimal field were obtained by Porter\cite{Porter}.
Recently Xiao and Niu have also studied the minimal field required
in the precessional magnetization switching in a conservative
system\cite{Xiao}. Their results were based on fast switching such
that the energy dissipation can be neglected and the magnetic
system can be regarded as a conservative system. They used the
phase plane to reveal the properties of magnetization reversal.
It was shown that precessional magnetic switching occurs when
localized trajectories in phase plane become delocalized.

Although it is reasonable to approximate a Stoner particle as a
conservative system in a short time if the damping (dissipation)
during its dynamical motion is small. A real magnetic particle
is not a truly conservative system when the damping is large
or if one is interested in the long time behave. Dissipation
should be taken into account and a Stoner particle should be
treated as a non-conservative system. In this paper we re-examine
the magnetization reversal as a nonlinear dynamical system in the
presence of dissipation. The magnetic dynamics of a single domain
magnetic particle can be described by the evolution trajectories
in the phase plane. One can use the general concepts of nonlinear
dynamics not only to understand all results from previous
studies, but also to see the validity conditions of some of these
results such as the SW-limit. The paper is organized as follows.
In Sec. II we first reformulate the magnetization reversal in
terms of nonlinear dynamics concepts, such as attractors and phase
flow. Previous results are re-interpreted in such language.
The Landau-Lifshitz-Gilbert (LLG) equation is introduced.
Our main results are presented in Sec. III. The conditions under
which the SW limit is valid are given. For a given magnetic
anisotropy, we shall show that there is a critical damping constant
above which the minimal switching field is the same as that of
SW-limit. The reason and meaning of this critical damping constant
are also given. The change of fixed points and their basins under a
magnetic field is investigated. We show that the field corresponding
to the ballistic reversal changes from the perpendicular direction
for a conservative system to a direction window in the
presence of energy dissipation. The conclusion is given in Section IV.

%------------------------------------------------------------%
\section{II. A nonlinear dynamics picture of magnetization reversal}

\subsection{II.1 Attractors, phase flow and magnetization reversal}

\begin{figure}[htbp]
 \begin{center}
\includegraphics[width=8.5cm, height=4.cm]{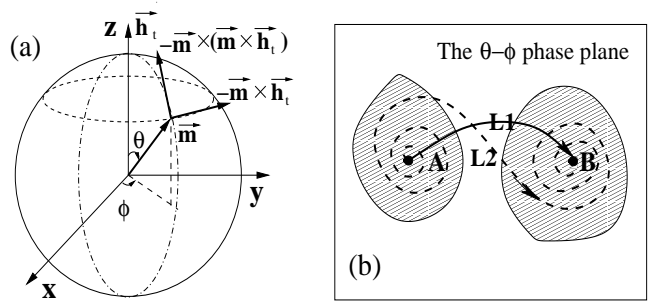}
 \end{center}
\caption{\label{fig1} (a) The magnetization $\vec{m}$ can be
uniquely determined by angles $\theta$ and $\phi$. $z$-axis
is assumed to be along the total magnetic field $\vec{h}_t$.
$-\vec{m}\times \vec{h}_t$ determines the precession direction,
and $-\vec{m}\times (\vec{m}\times \vec{h}_t)$ decides
the dissipation direction. (b) The $\theta-\phi$ phase plane
for the magnetization of a Stoner particle. Point A and point
B represent the initial and the target state, respectively.
Two shadowed areas denote schematically basins of two stable
fixed points A and B. The solid curve L1 and dashed curve L2
illustrate two different phase flow connected A and B. }
\end{figure}

Previous results on the magnetization reversal of Stoner particles
can be conveniently described in the terminology of nonlinear
dynamics. The phase space related to the magnetization is a
two dimensional (2D) plane because all atomic magnetic-moments are
aligned in the same direction, and the magnetization can rotate
under an external and/or an internal effective magnetic field that
can exert a torque on the spin. The polar angle $\theta$ and the
azimuthal angle $\phi$, shown in (Fig.~\ref{fig1}a), can fully
determine a magnetization $\vec{m}$. In the $\theta-\phi$ plane,
each point corresponds to a particular state of the magnetization.
A state will in general evolute to new states due to its dynamics.
Its motion can be described by a trajectory in the phase plane,
called phase flow. The phase flow for a dissipative system ends to
a few types of destiny (attractors), including fixed points, limit
cycles, or strange attractors. They correspond to stable states,
periodic, aperiodic and chaotic motions\cite{nbook}. In a 2D phase
plane, however, strange attractor solution is not allowed.
%It is convenient to use concepts of attractors and phase
%flow to investigate the dynamics of a
%system, especially those of low dimensional ones.

The only attractor related to the magnetization reversal of
Stoner particles is fixed points. The magnetization reversal
problem is as follows: Before applying an external magnetic
field, there are two stable fixed points (denoted by A and B
in Fig.~\ref{fig1}b), corresponding to magnetizations, say
$\vec{m}_0$ (point A) and $-\vec{m}_0$ (point B), along its
easy axis. The phase plane can be divided into two parts,
called basins of attractors. One is around A, and the other
around B, denoted by shadowed areas in Fig.~\ref{fig1}b.
The system in basin A(B) will end up at state A(B).
Initially, the magnetization is $\vec{m}_0$, and the goal is
to apply a small external field to switch the magnetization
to $-\vec{m}_0$ fast.

\subsection{II.2 Damping and non-damping magnetization reversals}

The conventional magnetization switching is based on a damping
mechanism through, typically, the spin-lattice relaxation. From
the point of view of nonlinear dynamics, the idea behind the
method is to construct the external magnetic field in such a way
that the target state is the only stable fixed point. In another
word, basin A (Fig.~\ref{fig1}b) is reduced to zero and the whole
(except probably a few isolated points) phase plane is the target
state basin (Basin B). The minimal reversal field (SW-limit) is
the one at which basin A shrinks to a point. Since the initial and
the target states have very large energy difference, the extra
energy must be dumped into the lattice during a spiral motion
before the system reaches the final state. The system first
spirals out of A, and then spirals toward B, denoted by phase flow
L2 in Fig.~\ref{fig1}b. This spiral motion is often
referred\cite{Miltat} to as ringing effect. The reversal time is
largely determined by the effectiveness of energy dissipation.

On the other hand, the target state does not need to be the only
fixed point in the recent fast magnetization switching. In fact,
it does not even need to be a fixed point. In the precessional
magnetization reversal, one applies a short magnetic field pulse
such that both initial and final states are not fixed points, and
system will start to flow in the phase plane. In order to switch
the magnetization, one needs to let the system to reach the basin
of the final state (Basin B) such that the system will flow to the
target state after the pulse field is switched off. Ideally, one
wants both initial and target states on its precessional path.
This is a non-damping method, and the reversal time does not rely
on the spin relaxation time. There are several ways with different
control precisions to move the system to the desired state. One
way is to apply a perpendicular pulse field to `kick' the system
to basin B. In comparison with conventional method, the spiral
motion out of the initial state is replaced by a
ballistic\cite{Miltat} motion. However, the system relies on
ringing effect to reach the final state. It was shown\cite{He2}
that the switching time can be reduced substantially, but it is
still hundreds of picoseconds for a normal magnetic particle due
to the ringing effect in the last stage of magnetization reversal.

The proposal of Xiao and Niu\cite{Xiao} is a fully ballistic
magnetization reversal scheme. Neglecting the energy dissipation
during the motion of a Stoner particle in a magnetic field,
the magnetization must move on an equal potential line.
Thus the idea is to apply an external magnetic field with a proper
strength and in a right direction such that both the initial and
the target states have the same energy and are on the same phase
flow trajectory as schematically illustrated by the solid line
L1 connected points A and B with an arrow in Fig.~\ref{fig1}b.
Without damping, there is no extra energy to dissipate in this
new approach, thus the system move from the initial state to the
final one in a ballistic way instead of in a ringing mode.
The typical time for a precession of 180$^{\circ}$ in a field of
teslas is order of picoseconds for usual magnetic materials so
a picoseconds magnetic field pulse is required in this method.
As soon as the system arrives at the target state, one needs
to switch off the external field, so the target state becomes
a fixed point of the system again. This approach thus requires a
precise control of the pulse duration.

From computational point of view, the magnetization reversal time
can be evaluated as soon as the phase flow connected the initial
and the target states is found. The reversal time is given by
the length $dl$ of phase flow line divided by the phase velocity
${\sqrt{\dot{\theta}^2+\dot{\phi}^2}}$ which is determined by the
system dynamics,
\begin{equation}
t=\int_A^B{\frac{dl(\theta, \phi)}
{\sqrt{\dot{\theta}^2+\dot{\phi}^2}}}.
\end{equation}
In the language of nonlinear dynamics, an external field modifies
the dynamics by changing the phase velocity field. This velocity
field is in general a continuous function of the external field.
A phase flow between the initial and target states could only
been set up when the external field is strong enough because the
initial and final states are two stable fixed points with equally
large basins at the beginning. The minimal switching field is
the critical one at which such a flow is created.

\subsection{II.3 The Landau-Lifshitz-Gilbert equation}

The magnetization dynamics of a Stoner particle is governed by the
Landau-Lifshitz-Gilbert (LLG) equation\cite{Landau},
\begin{equation*}
\frac{d\vec{M}}{dt'} = - |\gamma| \vec{M} \times \vec{H}_{t} +
\frac{\alpha}{M_s} \vec{M} \times \frac{d\vec {M}}{dt'},
\end{equation*}
which can also be written as
\begin{equation}
(1+\alpha^2)\frac{d\vec{M}}{dt'} = - |\gamma| \vec{M} \times
\vec{H}_{t} - \frac{\alpha |\gamma|}{M_s} \vec{M} \times
(\vec{M}\times \vec{H}_{t}).
\end{equation}
Here $|\gamma|=2.21 \times 10^5 (rad/s) /(A/m)$ is the
gyromagnetic ratio, $M_s$ is the saturated magnetization of
the particle, and $\alpha$ is the phenomenological
dimensionless damping constant. The typical experimental
values of $\alpha$\cite{Back} ranges from 0.037 to 0.22 for
different Co films.
The equation describes the motion of
magnetization vector $\vec{M}$ under an applied magnetic field
$\vec{H}$ in the presence of an arbitrary magnetic anisotropy
energy density function $W(\vec{M}, \vec{H})$. The total field
$\vec{H}_{t}=-\nabla_{\vec{M}} W(\vec{M}
,\vec{H})/\mu_0=\vec{H}_{eff}+\vec{H}$, where $\vec{H}_{eff}$
denotes the internal effective field due to the magnetic
anisotropy. $\mu_0=4\pi \times 10^{-7} N/A^2$ is the vacuum
magnetic permeability.

It is convenient to write the LLG equation in a dimensionless
form by defining $\vec{m}\equiv \vec{M}/M_s$, scaled field $\vec{h_t}
\equiv \vec{H_t}/M_s=\vec{h}_{eff}+\vec{h}$, scaled time $t \equiv
t'/(|\gamma|M_s)$, and $w(\vec{m}, \vec{h})\equiv W(\vec{M},
\vec{H})/(\mu_0 M_s^2)$. The equation for $\vec{m}$ becomes
%\begin{equation*}
%\frac{d\vec{m}}{dt} = - \vec{m} \times \vec{h}_{t} + \alpha
%\vec{m} \times \frac{d\vec{m}}{d\tau},
%\end{equation*}
%or
\begin{equation}
(1+\alpha^2)\frac{d\vec{m}}{dt} = - \vec{m} \times \vec{h}_{t}
- \alpha \vec{m} \times (\vec{m} \times \vec{h}_{t}). \label{LLG}
\end{equation}
As shown in Fig.~\ref{fig1}a, the first term in the right hand
side of Eq.~\eqref{LLG} describes the precession motion around the
total field and the second term decides the dissipation direction
toward the total field. In our analysis, we always use parameters
of Co as our references. For Co film\cite{Back}, $M_s=1.36\times
10^6 A/m$, thus the time unit is approximately $3.33ps$ and the
energy unit is $2.32 \times 10^6 J/m^3$.

In terms of $\theta-\phi$, the dynamics of the magnetization is
determined by the following non-conservative two-dimensional
nonlinear autonomous dynamical equations\cite{Miltat},
\begin{align}
(1+\alpha^2)\dot{\theta} &= -\alpha \frac{\partial w}{\partial
\theta} - \frac{1}{\sin\theta} \frac{\partial w}{\partial \phi},\nonumber\\
(1+\alpha^2)\dot{\phi} &= \frac{1}{\sin \theta} \frac{\partial
w}{\partial \theta} - \frac{\alpha}{\sin^2 \theta} \frac{\partial
w}{\partial \phi}.
\end{align}
Different particles is characterized by different magnetic
anisotropic energy density functions $w(\vec{m}, \vec{h})$.
In our analysis, we always assume it to be the uniaxial with
its easy axis along the x-direction. Due to the rotation
symmetry around the easy axis, the applied field can be
chosen in the x-z plane without losing generality.
The general form of $w(\vec{m}, \vec{h})$ can be written as
\begin{align}
w(\vec{m}, \vec{h}) &= f(m_x) - m_x h_x - m_z h_z,
\label{form}
\end{align}
where $h_x$ and $h_z$ are the applied magnetic field along x-
and z-axis, respectively.

This naive-looking nonlinear dynamical equation does not have
exact solutions yet. Analytical results can only be obtained in
some special cases. For example, analytical solutions were
found\cite{Kikuchi,Gillette} in the absence of the internal
effective field ($\vec{h}_{eff}=0$). The analytical analysis can
also be done if there is no energy dissipation as what is shown
in Reference \cite{Porter,Xiao}. In general, one can use the
fourth-order Runge-Kutta method to study the system numerically.
In this study, we will use following $f(m_x)$ in Eq.~\eqref{form}
with different $k_2$ and $k_4$, whenever we need to use numerical
results to illustrate our understandings,
\begin{equation}
f(m_x)  = - {1 \over 2}k_2 m_x^2 - {1 \over 4}k_4 m_x^4.
\label{model3}
\end{equation}
Here $k_2,\ k_4$, accounting for the strength of the anisotropy,
are positive numbers. In the next section, we shall present our
main findings.

%(1+\alpha^2)\sin \theta \dot{\phi} = h_z \sin \theta - h_x (\cos
%\theta \cos \phi + \alpha \sin \phi) \\
%&+ h_y (\alpha \cos \phi -\cos \theta \sin \phi) - K \sin \theta
%\cos \theta  \cos^2 \phi \\
%&- K \alpha \sin \theta \sin \phi \cos \phi,\\
%&(1+\alpha^2) \dot{\theta} = -\alpha h_z \sin \theta + h_x (\alpha
%\cos \theta \cos \phi - \sin \phi) \\
%&+ h_y (\alpha \cos \theta \sin \phi + \cos \phi) - K\sin \theta
%\sin \phi \cos \phi \\
%&+ \alpha K \sin \theta \cos \theta \cos^2 \phi.

\section{III. Results and Discussions}

\subsection{III.1 The exactness of SW-limit at infinitely large dissipation}

\begin{figure}[htbp]
 \begin{center}
\includegraphics[width=7.cm, height=5.5cm]{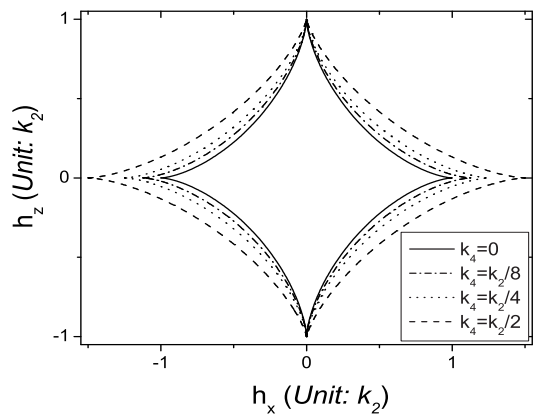}
 \end{center}
\caption{\label{fig2} The SW-limit for various choices of $k_2,\
k_4$. Solid curve: $k_4=0$; dash-dot curve: $k_4=k_2/8$; dotted
curve: $k_4=k_2/4$; dashed curve: $k_4=k_2/2$. }
\end{figure}

The conventional method is based on damping mechanism. Its
classical result is the so-called SW-limit. For an uniaxial model
with the easy axis along x-axis and magnetic field in xz-plane,
the SW-limit is obtained by assuming that the magnetization moves
in the xz-plane during its reversal. The minimal switching field
is given by\cite{Stoner}
%the one at which the only minimum energy state is at
\begin{align}
%\frac{\partial w}{\partial \theta}&=0,\\
%\frac{\partial^2 w}{\partial \theta^2}&=0.
\frac{d w}{d m_x }&=0,\\
\frac{d^2 w}{d m_x^2}&=0, \label{SW-curve}
\end{align}
with $m_x^2+m_z^2=1$. For a widely studied case of $k_2\neq 0$ and
$k_4=0$, the SW-limit\cite{Stoner} is
\begin{equation}
(h_x/k_2)^{2/3}+(h_z/k_2)^{2/3}=1,
\end{equation}
corresponding to the solid-line in Fig.~\ref{fig2}. The SW-limits
for various choices of $k_4$ are also plotted in Fig.~\ref{fig2}.

The original SW-limit was derived in the static case\cite{Stoner}.
As shown in the dynamical Eq.~\eqref{LLG}, the first term on the
right hand side will lift the magnetization away from the xz-plane.
Thus the assumption of the SW-limit that the magnetization moves
in the xz-plane is only true when this term can be neglected.
This will happen when damping constant becomes infinite
($\alpha \rightarrow \infty$). In this case, the magnetization
will move toward the total field as denoted by the big circle
passing through the north-south poles in Fig.~\ref{fig1}a.
This is the steepest energy descent path for the magnetization.
Thus, the minimal switching field of SW-limit corresponds to the
one at which there is only one minimum in energy landscape.

\subsection{III.2 Critical value of damping constant}

In a realistic system, as damping constant is not infinitely
large, the magnetization does not need to move along the steepest
energy descent path. As a result, a system may still move from
the initial state to the local minimum located near the target
state even when an external field is smaller than the SW-limit.
So, after the external field is removed, the system will move
toward the target state through a ringing mode, achieving the
magnetization switching. As it was shown in many previous
studies\cite{He1,Porter}, the minimal switching field can
be smaller than the SW-limit. Numerical calculations\cite{He1}
show that when the damping constant $\alpha<1$,
magnetization switching can occur well below the SW-limit.
While $\alpha\ge 1$, the minimal switching field is the SW-limit.
Thus, it implies a critical $\alpha_c$ exists, above which the
minimal switching field is given by the SW-limit.
In Reference 17, $\alpha_c=1$. To the best of our knowledge,
there is no clear understanding of this result. In fact,
previous studies\cite{He1} seem suggesting that $\alpha_c=1$
is very special. We want to show that there is indeed
a critical damping constant for a given magnetic anisotropy.
But, this critical value can be different for different
anisotropy, and $\alpha=1$ is not special.
We shall provide an explanation to the $\alpha_c$.

\begin{figure}[htbp]
 \begin{center}
\includegraphics[width=7.cm, height=6.cm]{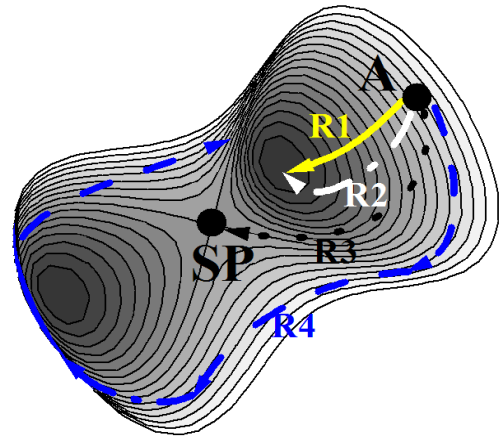}
 \end{center}
\caption{\label{fig3} (Color online) The schematic 3D energy
landscape plot of the case $h_1<h<h_{SW}$. Point A denotes the
initial state, whose energy is supposed to be higher than that of
the saddle point (SP). Lines R1, R2, R3 and R4 show schematically
four typical evolution trajectories for $\alpha=\infty$,
$>\alpha_i$, $\alpha_i$, and 0, respectively.}
\end{figure}

In order to understand the origin of the critical $\alpha$, let
us consider energy landscape under different external field.
As we mentioned in the previous section, there is only one
stable fixed point when $h>h_{SW}$. Asymptotically, the system
shall always end up at the fixed point for any non-zero damping.
Thus, if one switch off the field after it reaches the fixed
point, the system will surely move to the target state (state B).
There is also a
$h_1<h_{SW}$ at which the energy of system at the initial state
equals that at the saddle point between two stable fixed points.
Thus there is no way to switch the magnetization when $h<h_{1}$
because the energy of the initial state (A) is too smaller to
overcome the potential barrier between the initial and final
states. $h_1$ can be determined from the following equations
\begin{align}
\frac{d w}{d m_x}&=0,\label{saddle}\\
w(m_x) &=w_A,
\end{align}
where $w_A$ is the energy at the initial state ($m_x=1$). For a
field $h$ between $h_1$ and $h_{SW}$, $h_{1}<h<h_{SW}$, there
exist two stable fixed points with a saddle point in between.
Furthermore, the energy of the initial state is higher than that
of the saddle point. Fig.~\ref{fig3} is a schematic 3D plot of the
energy landscape for the case of $h_{1}<h<h_{SW}$. Point A denotes
the initial state, whose energy is supposed to be higher than that
of the saddle point (SP) between two local minima. In this case,
the flow starting from A will finally evolute into either of two
fixed points, depending on the value of $\alpha$. When $\alpha$ is
infinity, the system will evolute into the minimum near the
initial state along the steepest descent path, as shown by line
R1. For the opposite extreme of zero damping ($\alpha=0$), the
system will move along an equal potential contour (line R4)
surround the two minima (fixed points), as investigated in
Reference 19. For small $\alpha$, the magnetization can make many
turns around the two local minima before it falls into either one.
So there is a special $\alpha=\alpha_i$ with which the system just
touches the saddle point (SP) when it rolls down from A, denoted
by dotted line R3. For $\alpha>\alpha_i$, energy damping is too
strong for the system to `climb' over the saddle point, denoted by
line R2. Value $\alpha_i$ depends obviously on the magnetic field,
and critical damping constant $\alpha_c$ is the value of $\alpha_i$
at $h=h_{SW}$.

\begin{figure}[htbp]
 \begin{center}
\includegraphics[width=7.cm, height=5.5cm]{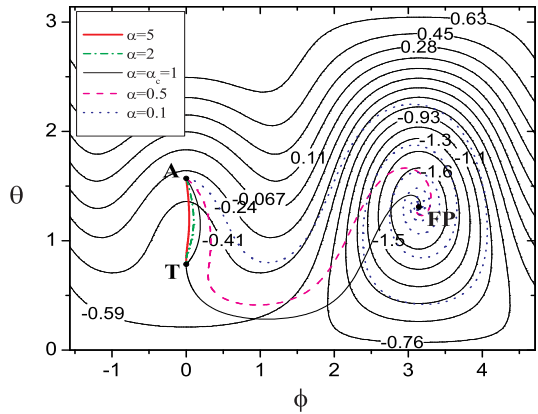}
 \end{center}
\caption{\label{fig4} (Color online) The contour plot of $w(\phi,
\theta)$ at $h=h_{SW}$ for the uniaxial magnetic anisotropy model
with $k_2=2$ and $k_4=0$. The field is along $3\pi/4$ to +x-axis.
Point A is the initial point. FP denotes the stable fixed point.
T denotes the inflexion point. All flow trajectories of $\alpha
\ge\alpha_c$ touch T while those with $\alpha < \alpha_c$ do not.}
\end{figure}

One may also understand the result from Fig.~\ref{fig4} of
trajectories of various $\alpha$ in the energy contour plot
at $h=h_{SW}$ along $135^{\circ}$ to +x-axis. The result in
the figure is for the uniaxial model with $k_2=2$ and $k_4=0$.
As mentioned early, the saddle point and one minimum merge
together at $h_{SW}$ to form an inflexion point denoted by T
in Fig.~\ref{fig4}. It is clear that all trajectories with
$\alpha>\alpha_c$ passing through the saddle-inflexion
point while all those with $\alpha<\alpha_c$ do not.
One may notice that all curves of $\alpha>\alpha_c$ do not
move after they reach point T because T is a saddle point.
But any small fluctuation will result in the system to
leave T and to end up in FP.

\begin{figure}[htbp]
 \begin{center}
\includegraphics[width=7.cm, height=5.5cm]{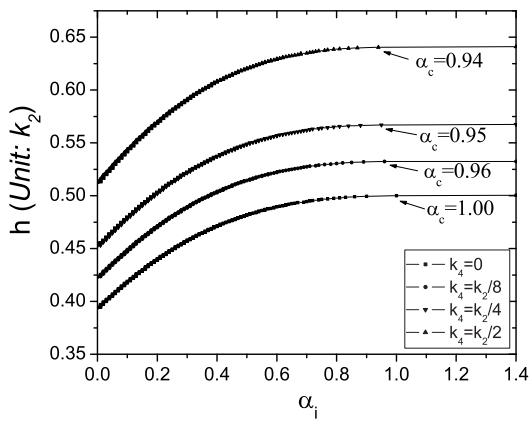}
 \end{center}
\caption{\label{fig5} The minimal switching field versus
damping constant $\alpha$. The field is along $3\pi/4$
to $+x$ axis. Square: $k_4/k_2=0$; Circle:
$k_4/k_2=1/8$; Down-triangle: $k_4/k_2=1/4$; Upper-triangle:
$k_4/k_2=1/2$. Smooth connection curves are to guide eyes.}
\end{figure}

In order to demonstrate the correctness of our reasoning of
existence of $\alpha_c$, and the value of the critical damping
constant varies with the magnetic anisotropy, we carried out
numerical calculations on the uniaxial magnetic anisotropy
model with different ratio of $k_4/k_2$. Fig.~\ref{fig5} is
$\alpha$-dependence of the minimal switching field for
$k_4/k_2=0;\ 1/8;\ 1/4;\ 1/2$, respectively. Indeed, all curves
(depend only on the ratio of $k_4/k_2$) saturate to their
corresponding SW-limit values $h_{SW}$ after $\alpha$ is greater
than certain values, the critical damping constants $\alpha_c$.
Furthermore, $\alpha_c$ is different for different $k_4$, varying
from $\alpha_c=1$ for $k_4=0$ to $\alpha_c=0.94$ for $k_4=k_2/2$.
Thus, $\alpha_c=1$ is not special at all!

\subsection{III.3 Changes of basins under an external field}

\begin{figure}[htbp]
 \begin{center}
\includegraphics[width=7.cm, height=5.5cm]{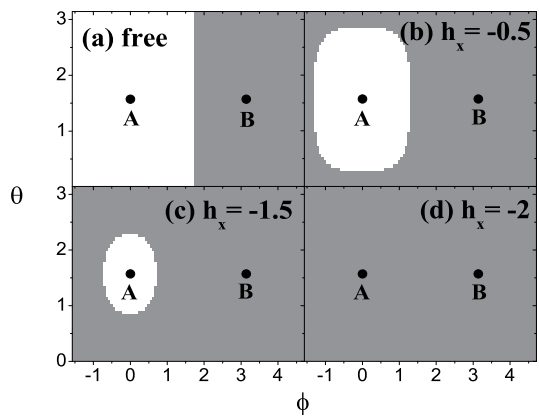}
 \end{center}
\caption{\label{fig6} The basins for $k_2=2$, $k_4=0$, and
$\alpha=0.1$ with various fields parallel to the easy axis. Points
A and B, which are also the fixed points, are the initial and the
target states. The white color is the basin of fixed point A and
the grey color is that of B. (a) In the absence of an external
field; (b) $h_x=-0.5$ ; (c) $h_x=-1.5$; (e) $h_x=-2$ (SW-limit).}
\end{figure}

\begin{figure}[htbp]
 \begin{center}
\includegraphics[width=7.cm, height=5.5cm]{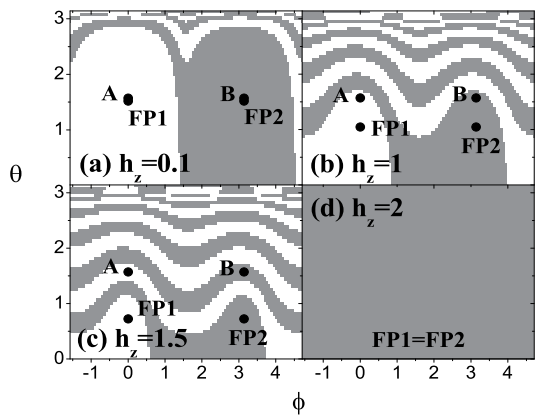}
 \end{center}
\caption{\label{fig7} The basins in the $\theta-\phi$ plane with
different strength of a field perpendicular to the easy axis, say
z-direction. Points A and B denote the initial and the target
states. The white color is the basin for fixed point FP1 and the
grey color is that for FP2. (a) $h_z=0.1$; (b) $h_z=1$; (c)
$h_z=1.5$; (d) $h_z=2$ (SW-limit). Other parameters are the same
as Fig.~\ref{fig6}. }
\end{figure}

\begin{figure}[htbp]
 \begin{center}
\includegraphics[width=7.cm, height=5.5cm]{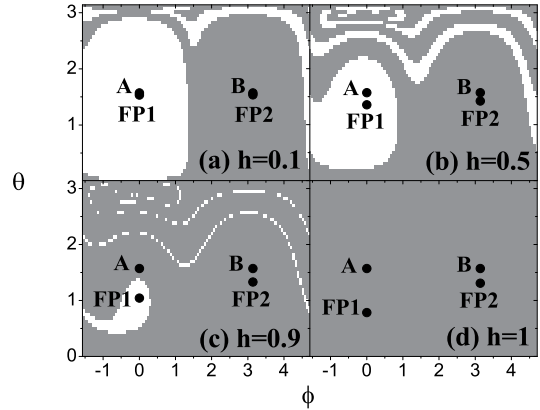}
 \end{center}
\caption{\label{fig8} The basins in $\theta-\phi$ plane
with different $h$ along $135^{\circ}$ to the +x axis.
Points A and B denote the initial and the target states.
The white color is the basin for fixed point FP1 and the
grey color is that for FP2. (a) $h=0.1$; (b) $h=0.5$;
(c) $h=0.9$; (d) $h=1$ (SW-limit). Other parameters are
the same as Fig.~\ref{fig6}. It demonstrates that A and B
could be connected by one phase flow trajectory in a field
with both non-zero x- and z-components.}
\end{figure}

We would like to investigate numerically how fixed points and
their basins change under a field along different directions. In
the following calculation, $k_2=2$, $k_4=0$, and $\alpha=0.1$ is
used. In the absence of an external magnetic field, there are two
fixed points at ($\theta=\pi/2, \phi=0$) and ($\theta=\pi/2,
\phi=\pi$), respectively. They are also the initial state (A) and
the target state (B). Depending on the direction and strength
of an applied field, two fixed points, FP1 and FP2, may move away
from A and B. The numerical procedures are as follows: Divide the
$\theta-\phi$ plane ($\theta\in (0,\pi)$, and $\phi\in (-\pi/2,
3\pi/2)$) into $100 \times 50$ meshes. Each mesh site represents
one particular state. Starting from an arbitrary state, the state
belongs to basin of FP1 (FP2) if the system evolutes into fixed
point FP1 (FP2) after a long enough time. Basin FP1 (FP2) is
colored white (grey). Fig.~\ref{fig6}-Fig.~\ref{fig8} are the
basins with different external fields.

{\bf A. Field along the easy axis}

Fig.~\ref{fig6} (a) is the basin without an external field. The
basins divide the phase plane into two equal-area parts with the
separatrix lines of $\phi=\pi/2$ and $3\pi/2$. When a field parallel
to the easy axis is applied, Points A and B are always two fixed
points of the system. However the areas of their basins vary.
Fig.~\ref{fig6} (b)-(d) show how the area of basin A shrinks while
that of B expands as the field increases. When the field reaches
the SW-limit (Fig.~\ref{fig6}d), the area of basin A shrinks to
zero. Thus B becomes the only stable fixed point and its basin
is the whole phase plane when the field is above the SW-limit.

{\bf B. Field perpendicular to the easy axis}

When a field perpendicular to the easy axis, say z-direction,
is applied, the variations of the fixed points and basins are
shown in Fig.~\ref{fig7}. The positions of two fixed points
shift along the $\theta$-axis, symmetrically located on
the two sides of +z-axis. It is evident from Fig.~\ref{fig7}
(a)-(c) that the basin shapes become layer-like structures.
The outer parts of the basins becomes layers first.
As the field increases, more basin areas become layers.
However, the total areas of two basins are equal due to the
symmetry of a perpendicular field when the field is below
the SW-limit. When the field is higher than a certain value,
point A will fall into the layered region (Fig.~\ref{fig7}c)
and may flow toward the right hand side of the phase plane.
Once the magnetization passes the separatrix line
($\phi=\pi/2$), it will evolute into the target state B
if one can switch off the field. This precessional procedure
was employed in experiments\cite{Back} and numerical
calculations\cite{He2}.

It is interesting to notice that the width of these layers become
generally thinner when they are away from their cores (central
parts of basins around fixed points), and the number of layers
increases with the field. The width of a layer is determined by
the size of basin core, energy variation in the phase plane, and
the energy dissipation during a $360^{\circ} $-precession ($\phi$
moves from $-\pi/2$ to $3\pi/2$). As the field increases, the
areas of the core basins shrink, and the precession period is
short as well. Thus layers become thinner while the layer number
increases. The energy dissipating rate is $\frac{dw}{dt}=-\frac
{\alpha}{1+\alpha^2}|\vec{m}\times\vec{h}_{t}|^2$. The precession
period can be estimated as about $(\vec{m}
\times\vec{h}_{t})^{-1}$ since $\vec{m} \times \vec{h}_{t}$ is the
angular frequency along the latitudes. So the energy loss is
proportional to $|\vec{m} \times \vec{h}_{t}| =|\vec
{h}_{t}\sin\theta|$, which is small for $\theta$ is near $\pi$.
Thus the width of the outer layers is thinner than that of inner
ones. Fig.~\ref{fig7}d shows when the field strength reaches
SW-limit, the two fixed points merge to the same point $\theta=0$
(north-pole).

{\bf C. Field at 135$^{\circ}$ to the easy axis}

We now investigate how the basins change under a field with
$135^{\circ}$ to the +x-direction since the minimal switching
field is smallest around this direction\cite{He1}.
Fig.~\ref{fig8} is the plot of fixed points and their basins
at different field strengths. It has both features of cases
with parallel and perpendicular fields. Increasing the field
strength, the fixed points shall shift along $\theta$-axis
while $\phi$ does not change. Unlike the case of perpendicular
field, the two fixed points have different $\theta$ values.
One can see that basin FP1 shrinks while that of FP2 expands
as the field strength increases, a feature with parallel field.
However, the layer structure also occurs in the outer parts of
the basins. The stripes become thinner when the field increases
and $\theta$ is away from the fixed point. At the SW-limit
(Fig.~\ref{fig8}d), the area of the basin of FP1 shrinks to zero.

\subsection{III.4 Ballistic path}

After seeing the change of fixed points and their basins under
different field in the previous section, let us investigate
the field and the reversal time for the ballistic connection.
Without dissipation, LLG equation is a conservative system.
A phase flow is an equipotential curve. As it was pointed out
in Reference\cite{Xiao}, only a perpendicular field is possible
to connect the initial and the target states ballistically.
Different from the conservative case\cite{Xiao}, the system
starting from A will never pass through the target state B in
the presence of dissipation. Even under an infinitely large
field, the energy loss during a $180^{\circ}$ precession is
not negligible. Although $180^{\circ}$ precession time $\tau$
decreases as inverse of magnetic field, $\tau\sim\pi(1+\alpha
^2)/h$ when $h>>1$ and $\alpha<<1$, the energy dissipation
rate goes as $\frac{dw}{dt}=-\frac{\alpha}{1+\alpha^2}
|\vec{m}\times\vec{h}_{t}|^2\propto h^{2}$, thus the energy
loss during $\tau$ is proportional to $h$ field\cite{note}!
In order to connect both A and B ballistically, one has to
create a small energy difference between A and B such that
the energy dissipated on its way from A to B equals the
energy difference.

On the other hand, Eq.~\eqref{LLG} can be solved exactly in the
absence of magnetic anisotropy ($k_2=0,\ k_4=0$)\cite{Kikuchi}
with solution $\phi=ht/(1+\alpha^2)$ and $\cos\theta=
[(1+\cos\theta_0)e^{2\alpha h t/(1+\alpha^2)}-1+\cos
\theta_0]/[(1+\cos \theta_0)e^{2\alpha h t/(1+\alpha^2)}+1
-\cos\theta_0]$, where $\theta_0$ is the initial angle between
the field and the magnetization (we assume the field is along
the z-axis and the initial $\phi$ is zero). Thus the ballistic
reversal corresponding to apply a field along direction $
\theta$ satisfying $-\cos\theta=[(1+\cos\theta)e^{2\alpha\pi}
-1+\cos\theta]/[(1+\cos\theta)e^{2\alpha \pi}+1-\cos\theta]$.
It is interesting to notice that the solution is unique and
the angle is $\tan (\theta/2)=e^{\alpha\pi/2}$.

\begin{figure}[htbp]
\begin{center}
\includegraphics[width=7.cm, height=5.5cm]{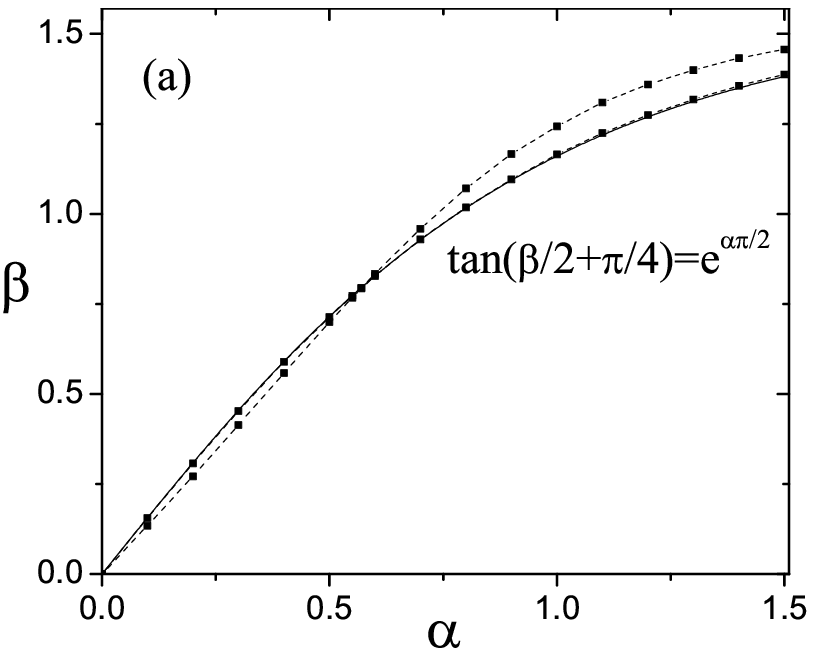}
\includegraphics[width=7.cm, height=5.5cm]{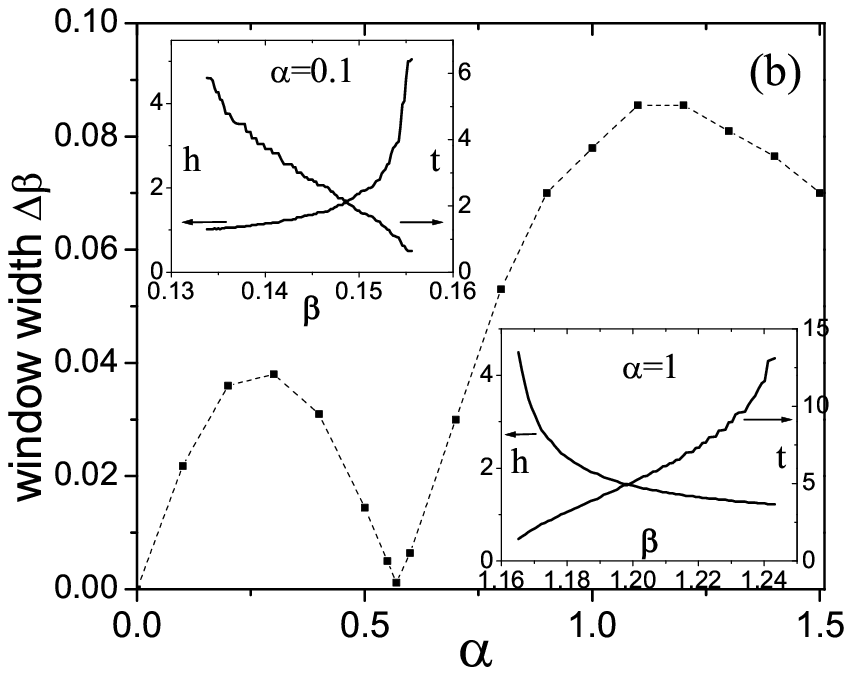}
 \end{center}
\caption{\label{fig9}(a) The upper and lower bounds of $\beta$ as
a function of damping constant $\alpha$. The solid line is $\tan
(\beta/2+\pi/4)=e^{\alpha \pi/2}$, which should be compared with
the upper bound of $\beta$ for $\alpha<0.57$ and the lower bound
for $\alpha>0.57$. The magnetic anisotropy is $k_2=2$ and $k_4=0$.
(b) The window width $\Delta \beta$ vs. $\alpha$ for $k_2=2$ and
$k_4=0$. Insets: the magnetic field and the corresponding reversal
time as a function of $\beta$ in the ballistic direction window.
$\beta \in (0.134,0.156)$ for $\alpha=0.1$ (upper-left inset),
$\beta \in (1.165,1.243)$ for $\alpha=1$ (lower-right inset).
The dashed lines are used to guide eyes.}
\end{figure}

Given a damping constant $\alpha$ and magnetic anisotropy,
described by $k_2$ and $k_4$, the $180^{\circ}$ precession
time $\tau(h,\beta)$ is a function of field strength $h
\equiv\sqrt{h_x^2+h_z^2}$ and its angle $\beta$ to the
z-axis ($\beta$ relates to $\theta$ by $\theta=\pi/2+\beta$).
Thus the energy dissipated $\Delta \epsilon(h,\beta)=\int_0
^{\tau}\frac{dw}{dt}dt$ during $\tau$ is also a function of
$h$ and $\beta$. For $h>>1$, one can neglect the magnetic
anisotropy, and above isotropic solution should be good.
Under the limit, $\Delta\epsilon$ is $2 h [1/(1+\tan^2(\beta
/2+\pi/4)e^{-2\alpha \pi})-1/(1+\tan^2(\beta/2+\pi/4))]$.
The energy difference $\Delta E(h,\beta)$ between A and B is
$2h\sin(\beta)$. Therefore, a ballistic path must satisfy
$\Delta\epsilon=\Delta E$ (a necessary condition but not a
sufficient one). Due to the symmetric reason, one needs to
consider only $\beta \in(0,\pi/2)$. Without energy
dissipation, the only solution is $\beta=0$ and any $h$
larger than certain minimal value. With large field ($h>>1$)
and energy dissipation ($\alpha\ne 0$), the approximate
solution is $\tan(\beta/2+\pi/4)=e^{\alpha\pi /2}$, the same
as the isotropic solution $\tan (\theta/2)= e^{\alpha\pi/2}$.
For $\alpha\ne 0$ and $k_2\ne 0$, we cannot solve $\Delta
\epsilon=\Delta E$ analytically. The field configuration of
the ballistic connection between A and B was found numerically.
The results were displayed in Figure \ref{fig9}.

Surprisingly, the field can be applied in a range of direction,
i.e. a direction window. Given a $\beta$ in this direction window,
$h$ is uniquely determined. Both the lower and the upper bound of
this $\beta$-window increase with the damping constant. Figure
\ref{fig9}a is the plot of the upper and the lower bounds of
$\beta$ as a function of $\alpha$. The solid line is  $\tan
(\beta/2+ \pi/4)=e^{\alpha\pi/2}$. Indeed, the window is around
the solid line. The width of the window depends both on the
damping constant and the magnetic anisotropy. At the zero and the
infinite damping constant, the width is zero. It is also zero in
the absence of magnetic anisotropy as illustrate by the exact
solution mentioned in the early paragraph. Thus, the width is
expected to oscillate with $\alpha$ for a given magnetic
anisotropy. This oscillation was indeed observed in numerical
calculations as shown in Fig. \ref{fig9}b for $k_2=2,\ k_4=0$. The
upper-left inset of Fig. \ref{fig9}b is the field and the
corresponding reversal time in the direction window for
$\alpha=0.1$ and $k_2=2,\ k_4=0$. In this particular case, $\beta$
is between 0.134 and 0.156. One sees that $h$ increases while
reversal time decrease with $\beta$. The similar plot for
$\alpha=1$ is shown in the lower-right inset of Fig. \ref{fig9}b.
Opposite to the case of small $\alpha(=0.1)$, $h$ decreases and
reversal time increases with $\beta$. Thus one should compare the
lower bound for $\alpha<0.57$ and the upper bound for
$\alpha>0.57$ with $\tan (\beta/2+\pi/4)=e^{\alpha \pi/2}$ since
it is expected to be exact for $h \rightarrow \infty$ when the
magnetic anisotropy can be neglected. An excellent agreement was
shown in Fig. \ref{fig9}a. Fig. \ref{fig9}b is the window width
$\Delta \beta$ as a function of $\alpha$. Our numerical results
indicate that the perpendicular configuration employed in the
current experiments\cite{Back,Schumacher} cannot achieve a fully
ballistic reversal. It should be pointed out that above results
are for the precise ballistic magnetization reversal. As we
mentioned early, other field can also switch magnetization if one
will also like to use the ringing effect at certain stages during
the reversal process.
%In fact, even the reversal time in Figure \ref{fig9}
%does not need to be the shortest.

\subsection{III.5 Discussions}

The SW theory is a classical work\cite{Stoner} which is based
on the energy consideration. Its connection to the LLG theory
was not discussed in literatures. We clarify that the SW theory
is the infinite dissipation limit of the LLG. Under the limit,
the magnetization moves along the steepest energy descent path.
The discrepancy between the SW-limit and the numerical minimal
switching field has been known for many years\cite{He1}.
Also, the existence of a critical value of $\alpha$ was known
numerically\cite{He1}. But its original was not given in the
previous theoretical studies\cite{He2,Bauer,He1,Porter,Xiao}.
To the best of our knowledge, its meaning is first revealed now.

So far we have reformulated the magnetization reversal in terms
of the language of nonlinear dynamics. The conventional reversal
technique is to make the target state to be the only fixed point
while the fast switching method is to connect the initial and the
target states on the same phase flow trajectory. Dissipation and
magnetic anisotropy play very interesting roles in a ballistic
magnetization reversal. Without dissipation, only perpendicular
fields larger than a minimal switching field can create a
ballistic path between the initial and the target states.
In the absence of a magnetic anisotropy and $\alpha\ne 0$, field
direction is not perpendicular but unique, and the field magnitude
is arbitrary. In the presence of both dissipation and anisotropy,
allowed directions for the ballistic connection form a window.
But inside the window, the field magnitude is single valued.
Due to the energy dissipation, applying a perpendicular field
to the easy axis of a uniaxial Stoner particle cannot directly
connect initial and final states by a phase flow trajectory.
Thus, the configuration employed in the current fast magnetization
reversal experiments\cite{Back,Schumacher} is not the best one.
A proper field along the direction window can improve the best
experimental numbers up to now.

There are advantages and disadvantages for the conventional and
the fast reversal methods. Although the switching time in the
conventional technique is at the nanoseconds, the technology is
less demanding and easy to implement. On the other hand, the fast
switching method could make the switching time at picoseconds
scale, but it needs to have precise control of the magnetic pulse.
If it is implemented, the cost should be much higher than that
of the conventional one. Besides time issue, minimal switching
field is another concern. The minimal switching field in the
conventional reversal technique occurs when only one energy
minimum exists in the basin of the final state. Depending on
the dissipation rate, the minimal field in the precessional
fast switching method may not be much smaller than the SW-limit.
So far, in both conventional and fast magnetization reversal
schemes, both the direction and the magnitude of the magnetic
field during the pulse duration are assumed to be fixed.
If the direction of the magnetic field is allowed to vary with
time, the minimal switching field can be even smaller.
However it will be more demanding to implement technologically.

%----------------------------------------------------------------------------
\section{IV. Conclusion}

In conclusion, we show that the magnetization reversal can be
conveniently examined in the terminology of nonlinear dynamics.
The presence of the dissipation will not fail the fast switching
method. We clarify that the SW result is the limit of LLG theory
with an infinitely large dissipation. In this limit, the
magnetization moves along the steepest energy descent path. We
show that there is a critical value of the damping constant for a
given magnetic anisotropy, above which, the minimal switching
field is the same as the prediction of SW theory. Based on the
change of fixed points and their basins under an external field in
different directions, we show that the magnetization reversal time
can be much shorter when the field drives both initial and target
states away from fixed points, but puts them on the same phase
flow trajectory. In the absence of energy dissipation, the field
should be applied perpendicularly to the easy axis in order to
achieve a ballistic magnetization reversal. However, with both
energy dissipation and a magnetic anisotropy, the field can be
applied along a direction window. The width of the window depends
on both the damping constant and the magnetic anisotropy. It is
zero for either zero damping constant or zero magnetic anisotropy
($k_2=0$ and $k_4=0$). Unlike the conventional magnetization
reversal method, the new scheme demands a precise control of
picoseconds pulse of a magnetic field.

%----------------------------------------------------------------------------
\section{Acknowledgments}
We would like to thank D. Xiao and Q. Niu for sending us their
preprint before submitting for publication.
This work is supported by UGC, Hong Kong, through RGC CERG grants.


\begin{thebibliography}{}
\bibitem{note} One cannot use a small $h$ to switch a
magnetization in the presence of a non-zero magnetic anisotropy.
A finite $h$ is needed in order to overcome the potential
barrier along a reversal path.
\end{thebibliography}
\end{document}